\documentclass[useAMS,usenatbib,usegraphicx]{mn2e}

\title[Alfv\'{e}n Wave Amplification]{Alfv\'{e}n Wave Amplification 
and Self-Containment of Cosmic-Rays
Escaping from a Supernova Remnant}
\author[Y. Fujita et al.]{Yutaka Fujita$^{1}$\thanks{E-mail:
fujita@vega.ess.sci.osaka-u.ac.jp (YF)}, Fumio Takahara
$^{1}$, Yutaka Ohira$^{2}$, and Kazunari Iwasaki$^{3}$
\\
$^{1}$Department of Earth and Space Science, Graduate School
of Science, Osaka University, \\
1-1 Machikaneyama-cho, Toyonaka, Osaka
560-0043, Japan\\
$^{2}$Theory Centre, Institute of Particle and Nuclear Studies, 
KEK, 1-1 Oho, Tsukuba 305-0801, Japan\\
$^{3}$Department of Physics, Nagoya University, Nagoya
464-8602, Japan
}
\begin{document}

\date{Accepted 0000 December 00. Received 0000 December 00; in original
form 0000 October 00}

\pagerange{000--000} \pubyear{0000}

\maketitle

\label{firstpage}

\begin{abstract}
 We study the escape of cosmic-ray (CR) protons accelerated at a
 supernova remnant (SNR) by numerically solving a diffusion-convection
 equation from the vicinity of the shock front to the region far away
 from the front. We consider the amplifications of Alfv\'{e}n waves
 generated by the escaping CR particles and their effects on CR escape
 into interstellar medium (ISM). We find that the amplification of the
 waves significantly delays the escape of the particles even far away
 from the shock front (on a scale of the SNR). This means that the
 energy spectrum of CR particles measured through $\gamma$-ray
 observations at molecular clouds around SNRs is seriously affected by
 the particle scattering by the waves.
\end{abstract}


\section{Introduction}
\label{sec:intro}

Cosmic rays (CRs) are relativistic charged particles and most of them
are believed to be accelerated at the shock front of supernova remnants
(SNRs). In fact, TeV $\gamma$-ray emissions have been detected with
H.E.S.S. from the shell of young SNRs \citep{aha04,aha05}. Moreover, GeV
$\gamma$-ray emissions have been detected with Fermi and AGILE from
middle-aged SNRs surrounded by molecular clouds. They are likely to be
generated via a hadronic process, i.e. inelastic collisions between CR
protons accelerated at SNRs and ambient protons
\citep{abd09a,abd10a,abd10b,abd10c,tav10,giu10}. TeV $\gamma$-ray
emissions have also been discovered from some of these clouds
\citep{aha08}. The energy spectrum of CR protons is not described by a
single power-law. Thus, it is most likely to be affected by an escape
process from SNR shocks \citep[but see also][]{uch10}.

The $\gamma$-ray observations of molecular clouds around SNRs have given
us information not only on the CR particle acceleration in the SNRs but
also on the escape of the particles from the SNRs
\citep[e.g.][]{ptu03,cap09,ohi10,kaw11,ohi11}. In particular, their
escape into the surrounding interstellar medium (ISM) is important to
compare the $\gamma$-ray observations of molecular clouds around the
SNRs with theoretical models of particle acceleration. Recently,
\citet{lee08} and \citet{gab09} theoretically studied the escape into
the ISM and obtained an energy spectrum of the escaped particles. In
their studies, they assumed that the diffusion coefficient for the
particle diffusion outside the SNR is the one in the general region in
the Galaxy. However, \citet{fuj09a} suggested that the observations seem
to show that the diffusion time-scale of CRs around SNRs is much longer
than that in the general region in the Galaxy. They estimated that the
diffusion coefficient must be less than $\sim 1$~\% of that in the
general region to be consistent with the observations of the SNR W~28
and another possible SNR in Westerlund~2 \citep[see
also][]{tor08}. Motived by these results, \citet{fuj10} theoretically
studied the diffusion of particles in ISM after they left the shock
neighbourhood. They used Monte-Carlo simulations to follow a particle
motion, while at the same time they calculated the evolution of the
SNR. They considered the excitation of Alfv\'en waves through the
streaming instability caused by CRs, and the interactions between the
waves and the particles. Since the growth of the waves changes the
diffusion coefficient, they took the evolution of the diffusion
coefficient into consideration. They found that the particles actually
excite Alfv\'en waves around the SNR on a spatial scale of the SNR
itself if the ISM is highly ionised. Thus, even if the particles can
leave the shock neighbourhood, scattering by the waves prevents them
from moving further away from the SNR. This means that the particles
cannot virtually escape from the SNR until a fairly late stage of the
SNR evolution.

However, the model developed by \citet{fuj10} may still be rather
simplified. In particular, they separately treated the acceleration of
particles at the shock front and their propagation into the surrounding
ISM. Since both of the processes follow the same transport equation,
they should be treated seamlessly. In this paper, we present the results
of our numerical simulations that deal with both the particle
acceleration and the diffusion into the ISM at the same time. In these
simulations, we take account of the streaming instability developed by
the flux of CRs in order to study how the growth of the Alfv\'en waves
affects the diffusion of the CRs. We emphasise that we focus on the
propagation of CRs far away from the shock front ($r-R_s\sim R_s$, where
$r$ is the distance from the SNR centre and $R_s$ is the shock radius),
in contrast with previous studies that treat the escape around the shock
front \citep[$r-R_s\ll R_s$; e.g.][]{vla06}. As far as we know, this is
the first time to solve a transport equation from the shock vicinity to
the region far away from the shock front.

\section{Models}

\subsection{Equations}

The diffusion of CR particles obeys the transport equation
(diffusion-convection equation):
\begin{equation}
\label{eq:trans}
 \frac{\partial f}{\partial t} = \nabla(\kappa\nabla f) 
- \mbox{\boldmath{$w$}}\nabla f + \frac{\nabla\mbox{\boldmath{$w$}}}{3}
p\frac{\partial f}{\partial p} + Q \:,
\end{equation}
where $f(r,p,t)$ is the distribution function of the particles, $p$ is
the momentum, $\kappa$ is the diffusion coefficient,
\mbox{\boldmath{$w$}} is the hydrodynamic velocity of the background
gas, and $Q=Q_0\delta(r-R_s)$ is the source term for the particles,
which are injected at the shock front ($r=R_s$) at a momentum of $p_{\rm
inj}$. The coefficient $Q_0$ is given by
\begin{equation}
 Q_0=\epsilon\frac{\rho_1 u_1}{m}
\frac{\delta(p-p_{\rm inj})}{4\pi p_{\rm inj}^2}\:,
\end{equation}
where $\rho_1$ and $u_1$ are the gas density and the gas velocity
relative to the shock just upstream the shock front, respectively. In
\S~\ref{sec:results}, we set $p_{\rm inj}=2\: m c_{s2}$, where $c_{s2}$
is the sound velocity behind the shock front. The fraction of gas
particles that go into the acceleration process at the shock front is
$\epsilon=10^{-4}$.

The equations for the background gas are
\begin{equation}
 \frac{\partial\rho}{\partial t} + \nabla(\rho\mbox{\boldmath{$w$}})
= 0\:,
\end{equation}
\begin{equation}
 \rho\frac{\partial\mbox{\boldmath{$w$}}}{\partial t}
+ \rho(\mbox{\boldmath{$w$}}\cdot\nabla)\mbox{\boldmath{$w$}}
= - \nabla(P_c + P_g)\:,
\end{equation}
\begin{equation}
\label{eq:Pg}
 \frac{\partial P_g}{\partial t} + (\mbox{\boldmath{$w$}}\cdot\nabla)P_g
+ \gamma_g (\nabla\mbox{\boldmath{$w$}}) P_g = 0\:,
\end{equation}
where $\rho$, $\gamma_g$, and $P_g$ are the density, specific heat
capacity ratio, and pressure of the gas. The CR pressure is given by
\begin{equation}
 P_c = \frac{4\pi c}{3}\int_{p_{\rm min}}^\infty dp 
\frac{p^4 f}{\sqrt{p^2 + m^2 c^2}} \:,
\end{equation}
where $p_{\rm min}$ is the minimum momentum of injected CR particles,
$m$ is the mass of the particles, and $c$ is the velocity of light. We
consider only protons and neglect electrons.

\subsection{Diffusion Coefficient and Initial Conditions}

Since we treat the propagation of CRs away from the shock front, we
cannot assume a plane geometry. Therefore, we assume spherical symmetry
and solve equations (\ref{eq:trans})--(\ref{eq:Pg}) mainly based on the
numerical methods developed by \citet{ber94}. We do not fix the
diffusion coefficient $\kappa$ at the value corresponding to the Bohm
diffusion, which is different from the studies by
\citet{ber94}. Instead, we follow the evolution of $\kappa$ in this
study, considering the amplification of Alfv\'{e}n waves by CR
particles. Since the waves scatter the particles, they affect the
diffusion coefficient in equation (\ref{eq:trans}). The growth of the
waves is given by
\begin{equation}
\label{eq:psi}
 \frac{\partial \psi}{\partial t} 
\approx \frac{4\pi}{3}\frac{v_{\rm A} p^4 v}{U_{\rm M}}
|\nabla f|\:,
\end{equation}
where $\psi(t,\mbox{\boldmath{$r$}}, p)$ is the energy density of
Alfv\'en waves per unit logarithmic bandwidth (which are resonant with
particles of momentum $p$) relative to the ambient magnetic energy
density $U_{\rm M}$, and $v_{\rm A}$ is the Alfv\'en velocity
\citep{ski75,bel78}. We do not consider the damping of the waves for
simplicity. The diffusion coefficient is
\begin{equation}
\label{eq:kappa}
 \kappa=\frac{4}{3\pi}\frac{p v c}{e B_0 \psi}\frac{\rho_0}{\rho}\:,
\end{equation}
where $v$ is the velocity of the particle, $e$ is the elementary charge,
$B_0$ is the unperturbed magnetic field, and $\rho_0$ is the density of
unperturbed ISM. Following \citet{ber94}, we included the influence of
gas compression on the magnetic field ($B=B_0\rho/\rho_0$) and particle
scattering by assuming that $\kappa$ depends on the gas density.

The diffusion coefficient corresponding to the Bohm diffusion is given
by $\kappa_{\rm B0}=\rho_{\rm B} c/3$, where $\rho_{\rm B}=pc/(e B_0)$
is the gyroradius of particles. We calculate the evolution of the gas
and CR particles using the methods of \citet{ber94} for $r\la
r_b(p) \equiv 4\kappa_{\rm B0}/V_s + R_s$ and $p/(m c) \ga
10^{-3.4}$, where $V_s$ is the shock velocity. We call this region the
``shock region''. Since we are also interested in the CR diffusion
outside this region, we added equal interval meshes for $r_b(p)\la
r\la r_{\rm out}(p) \equiv 500\:\kappa_{\rm B0}/V_s + r_{\rm
b}$. We call this surrounding region the ``escape region''. For this
region, we calculate the diffusion of the CR particles only with $p/(m
c) \ga 10^{2.4}$. For this energy range, the escape region is
sufficiently apart from the shock front, and the feedback of the CR
particles on hydrodynamics is small there. Thus, we ignore the feedback
in this region ($\nabla \mbox{\boldmath{$w$}}\sim 0$) and solve
equation~(\ref{eq:trans}) without considering the change of $f$ in the
direction $p$ (the third term in the right hand of
equation~[\ref{eq:trans}]), which makes calculations much easier. The
term $\mbox{\boldmath{$w$}}\nabla f$ in equation~(\ref{eq:trans}) can
also be ignored in this region, although we do not omit the term in the
calculations. On the other hand, we do not solve the diffusion of
particles with $p/(m c) \la 10^{2.4}$ in the escape region, because
$\nabla \mbox{\boldmath{$w$}}$ cannot be ignored even in the
region. Since the feedback of the low-energy particles on hydrodynamics
is limited in the vicinity of the shock, they do not affect the
following results at $r-R_s\sim R_s$. We set $f=0$ at $r=r_{\rm out}$.

The initial condition of $\psi$ is given as follows. The diffusion
coefficient of the background ISM is given by
\begin{equation}
\label{eq:kappa_ism}
 \kappa_{\rm ISM}=10^{28}{\rm\: cm^2\: s^{-1}}
\left(\frac{E}{10\rm\: GeV}\right)^{0.5}
\left(\frac{B_0}{3\:\mu\rm G}\right)^{-0.5}
\:,
\end{equation}
where $E$ is the particle energy \citep{gab09}. We relate $\kappa_{\rm
ISM}$ to the energy density of Alfv\'en waves ($\psi_{\rm ISM}$):
\begin{equation}
\label{eq:psi_ism}
 \kappa_{\rm ISM}=\frac{4}{3\pi}\frac{p v c}{e B_0 \psi_{\rm ISM}}
\:.
\end{equation}
If we set $\psi=\psi_{\rm ISM}$ at $t=t_{\rm inj}$, where $t_{\rm inj}$
is the time when we start injecting CRs, $\psi$ does not grow enough and
particles are not accelerated.  As has been well known, for diffusive
shock acceleration to work, $\psi$ must be as large as that for Bohm
diffusion near the shock front \citep[e.g.][]{bel78,luc00}. Thus, in
this study, we simply assume that the diffusion at the shock front can
be represented by Bohm diffusion ($\kappa=\kappa_{\rm
B0}\rho_0/\rho$). One can find that the energy density of Alfv\'en waves
corresponding to $\kappa_{\rm B0}$ is $\psi_{\rm B}=4v/(\pi c)\approx
4/\pi$.  On the other hand, for $r-R_s \gg \kappa_{B0}/V_s$, the
coefficient should be the one at the interstellar space
($\kappa_0=\kappa_{\rm ISM}$). As the initial condition of $\psi$ at
$t=t_{\rm inj}$, we interpolate the two $\psi$s:
\begin{eqnarray}
\label{eq:psi_i}
 \psi_i(r,p) &=& (\psi_{\rm B} - \psi_{\rm ISM}) 
\exp\left[-\frac{(r-R_s(t_{\rm inj}))V_s(t_{\rm inj})}
{a_i\kappa_{\rm B0}}\right]\nonumber \\
& & + \psi_{\rm ISM} \:,
\end{eqnarray}
where $a_i$ is a parameter. Although the value of $a_i$ cannot be
specified, it would not be much larger than one. Therefore we assume
$a_i=5$ in the following simulations. The larger $a_i$ gives the smaller
diffusion coefficient at a given $r$ and results in less escape of the
particles. For values of $a_i$ as small as one, it is increasingly
difficult to obtain stable and precise results, because $f$ changes
rapidly at the boundary between the shock and escape region, where
resolution is not high enough. We consider three models: (A) the growth
of $\psi$ is considered, (B) the diffusion coefficient is fixed at
$\kappa=\kappa_{\rm B0}\rho_0/\rho$ regardless of time and position, and
(C) the wave energy density does not change and is fixed at
$\psi=\psi_i$, replacing $t_{\rm inj}$ with $t$ in
equation~(\ref{eq:psi_i}). The models B and C are calculated for
comparison.  In Model~A, we set a lower limit of $\psi$ as
$\psi\ge\psi_i$, replacing $t_{\rm inj}$ with $t$ in
equation~(\ref{eq:psi_i}). Moreover, we set a upper limit of $\psi$ as
$\psi\le\psi_{\rm B}$.

Since we do not consider the neutral damping of the waves, we choose the
ISM with a relatively high temperature. The density and sound velocity
of the background ISM is $\rho_0=7.0\times 10^{-27}\rm\: g\: cm^{-3}$
and $c_s=154\rm\: km\: s^{-1}$, respectively. A supernova explodes at
$t=0$ and $r=0$ with an energy of $10^{51}$~erg. The background magnetic
field is $B_0=\sqrt{8\pi U_M}=3\:\mu\rm G$. We inject CR particles and
calculate the growth of Alfv\'en waves (equation~[\ref{eq:psi}]) for
$t>t_{\rm inj}=0.55\: t_0$, where $t_0$ is the time when the free
expansion phase of the SNR ends and the Sedov phase begins. For the
parameters we adopted, $t_0=4.7\times 10^3$~yr.

\section{Results and Discussion}
\label{sec:results}

\begin{figure}
\includegraphics[width=84mm]{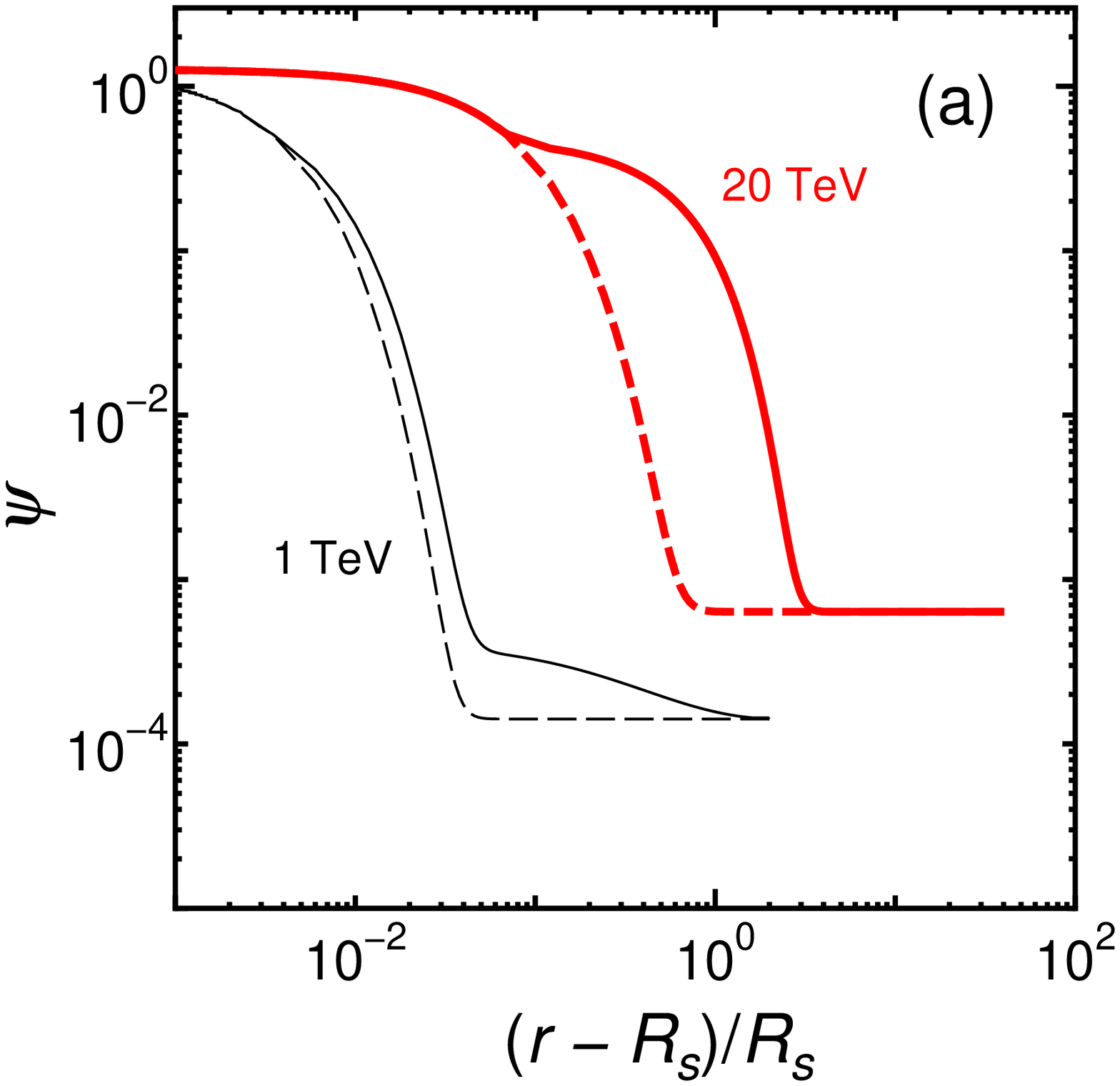}
\includegraphics[width=84mm]{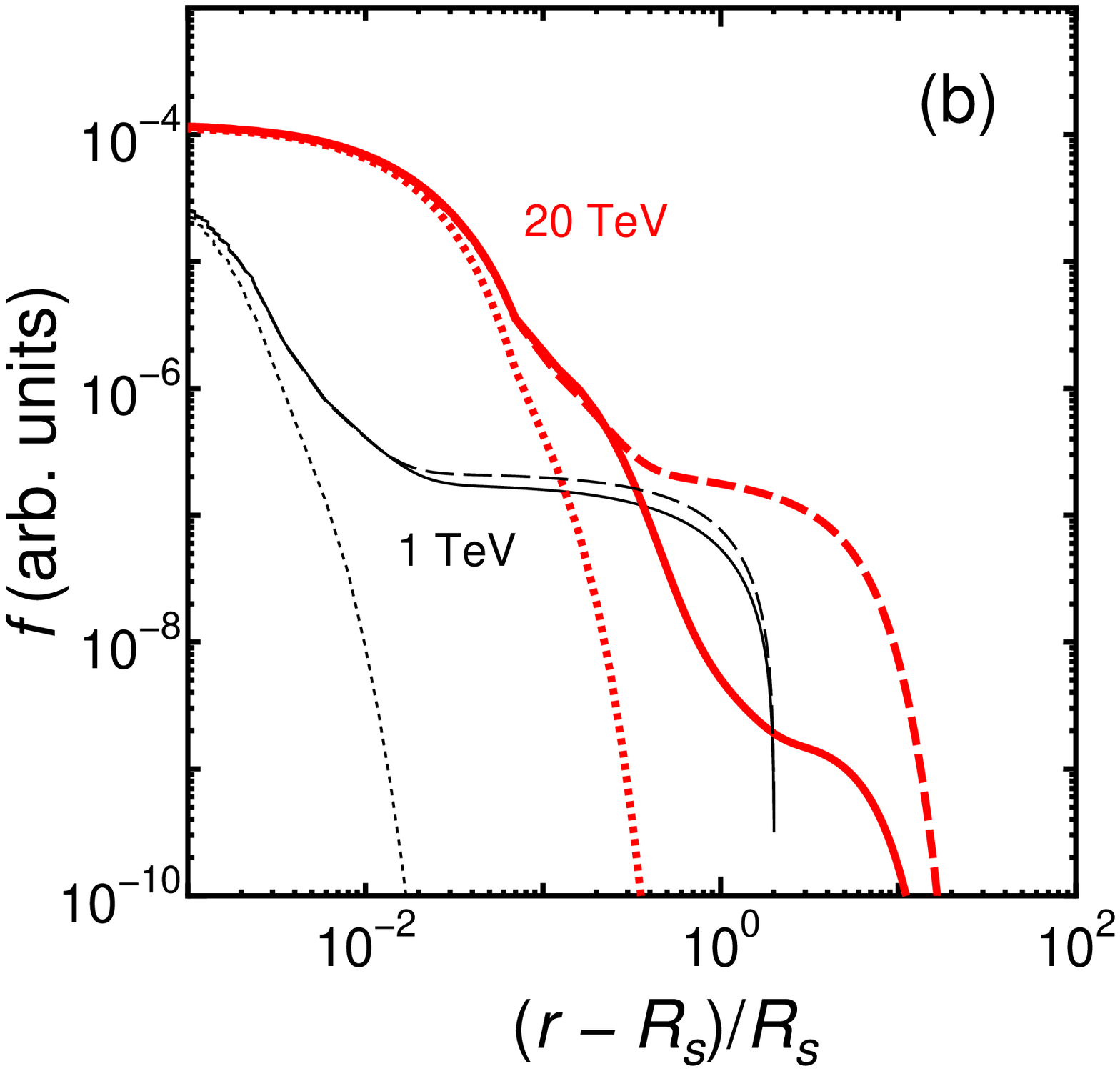}
\caption{Profiles of wave
energy density $\psi$ at $t=10\: t_0$. Thin lines correspond to waves
interacting with particles with $pc=1$~TeV and thick lines correspond to
those with $pc=20$~TeV. Solid lines are for Model~A and dashed lines are
for Model~C. (b) Same as (a) but for CR distribution
$f$.\label{fig:Bf}}
\end{figure}

In Figure~\ref{fig:Bf}a, we show the profiles of $\psi$ at $t=10\: t_0$
for Models~A and C. At this time, the radius of the SNR is $R_s=73$~pc,
and the compression ratio of the gas is $\rho_2/\rho_0=3.6$, where
$\rho_2$ is the gas density behind the shock, for Models~A, B,
and~C. The left ends of the curves correspond to $\psi_{\rm B}$, and the
right ends correspond to $\psi_{\rm ISM}$. The difference between
Model~A and~C indicates that CR particles can significantly amplify
Alfv\'en waves. In particular, the amplification occurs even in the
region far away from the shock front (say, at $r\sim 2\: R_s$) for
particles with $pc=20$~TeV. The profiles are given by the balance
between the growth through streaming instability and the flow of
low-$\psi$ gas from upstream on the shock coordinate. The amplification
of the waves prohibits the diffusion of particles. In
Figure~\ref{fig:Bf}b, we show the profiles of $f$ at $10\: t_0$. While
the particles are confined around the shock front in Model~B, they can
extend to the region where $\psi\sim \psi_{\rm ISM}$ in Model~C
(Figure~\ref{fig:Bf}). The results when the growth of $\psi$ is
considered (Model~A) are located between those two models, although the
profiles of $\psi$ and $f$ are somewhat different between $pc=1$ and
20~TeV (Figure~\ref{fig:Bf}). The overall features of $\psi$ and $f$ do
not change for $t\ga t_0$.

\begin{figure}
\includegraphics[width=84mm]{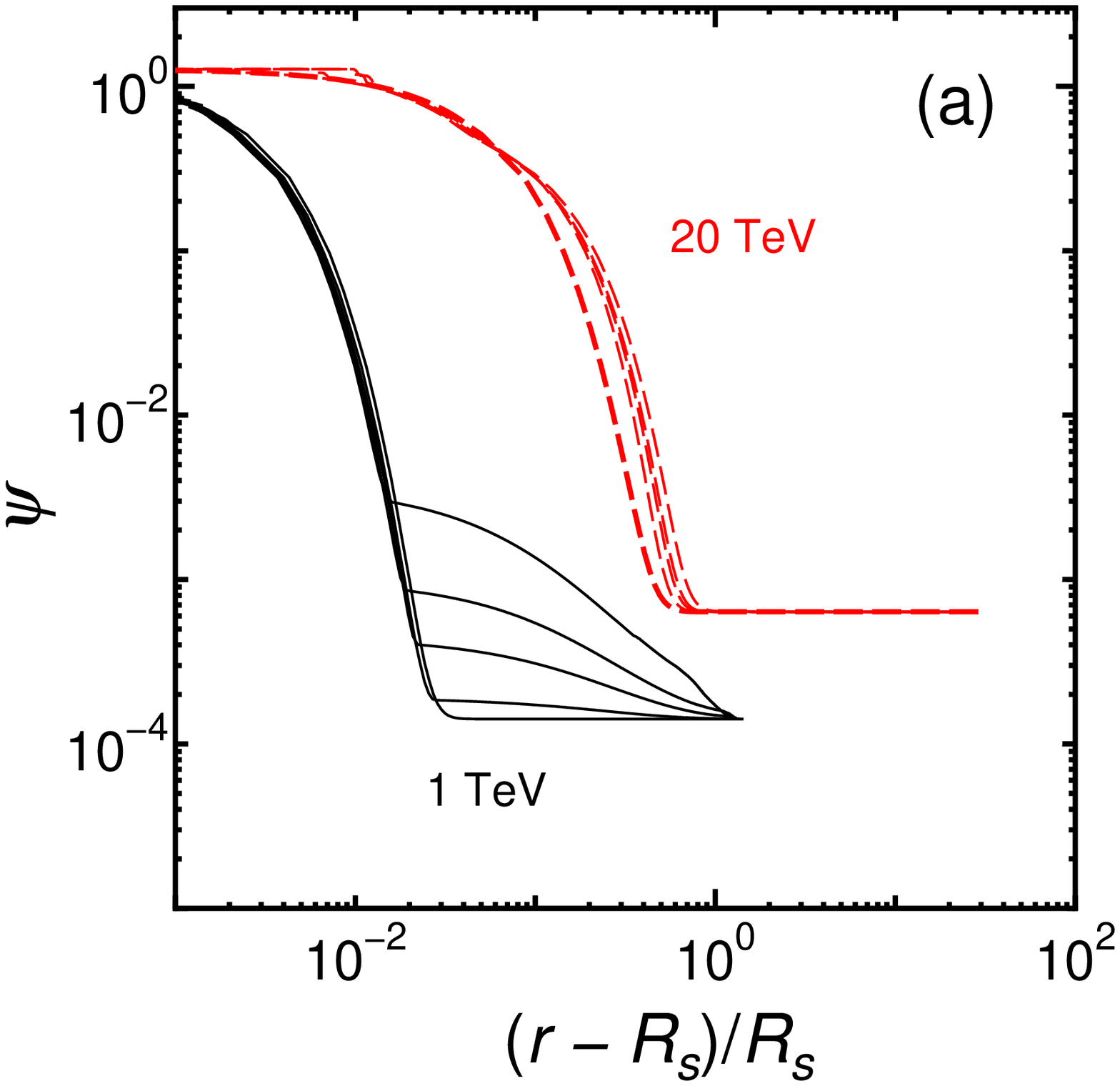}
\includegraphics[width=84mm]{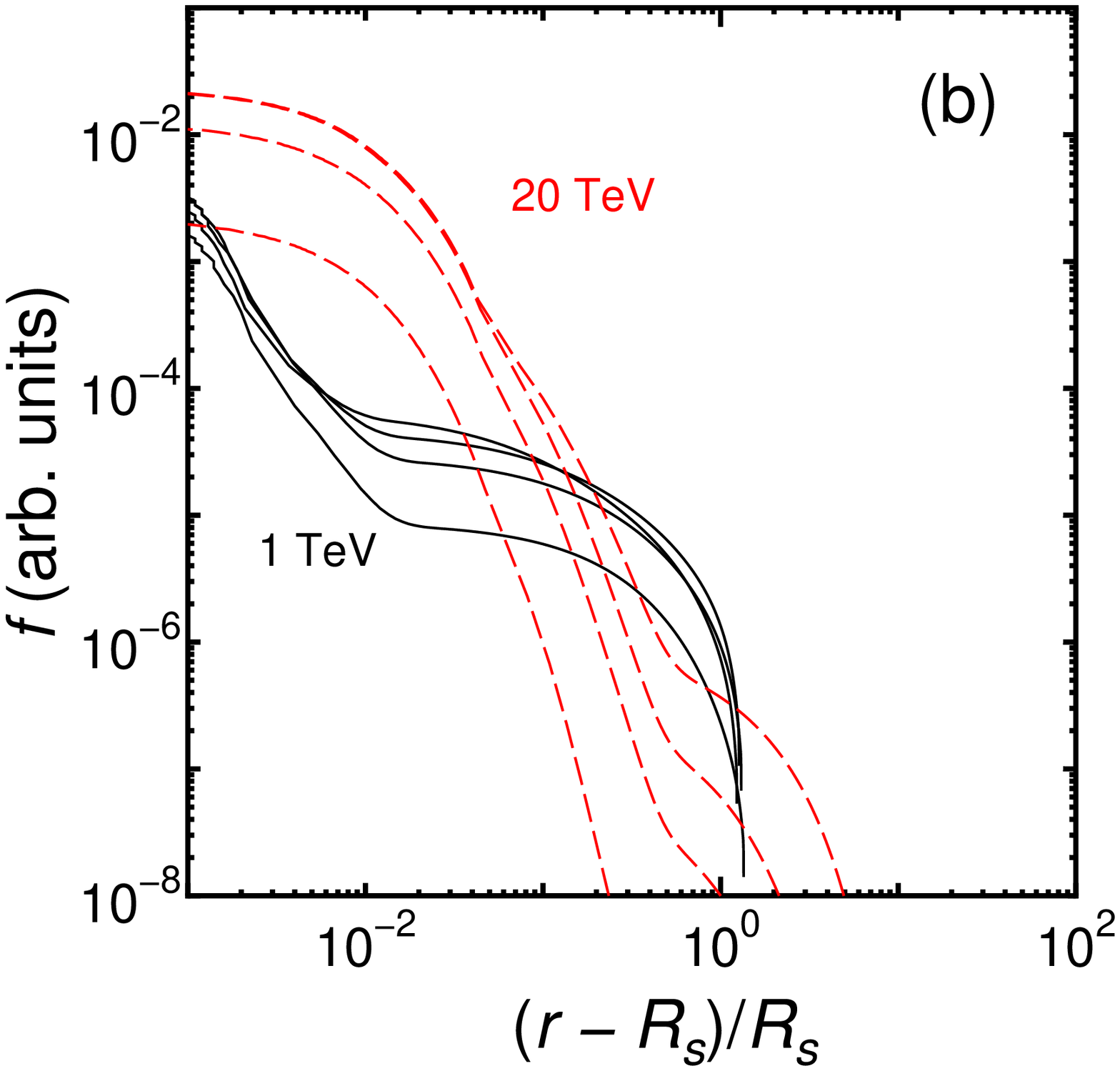}
\caption{(a) Profiles of wave
energy density $\psi$ for Model~A at $t=0.55\: t_0$, $0.60\: t_0$,
$0.65\: t_0$, $0.70\: t_0$, and $1.0\: t_0$ (from bottom to top). (b)
Same as (a) but for CR distribution $f$ at $t=0.60\: t_0$, $0.65\: t_0$,
$0.70\: t_0$, and $1.0\: t_0$ (from bottom to top).\label{fig:evo}}
\end{figure}

The results for Model~A can be explained as follows.
Figure~\ref{fig:evo} shows the evolution of $\psi$ and $f$ for Model~A
for $t_{\rm inj}<t\leq t_0$, normalised by $R_s(t)$ in the radial
direction. The radius of the SNR at $t=t_0$ is $R_s=23$~pc. For
$pc=1$~TeV, $\psi$ increases in the ``instep'' or the region where the
gradient of $\psi$ is small ($0.02\la (r-R_s)/R_s\la 2$). This is
because of the leak of particles into the low-$\psi$ region ($\psi\sim
\psi_{\rm ISM}$). On the other hand, for $pc=20$~TeV, $\psi$ increases
in the ``shin'' or the region where the gradient of $\psi$ is large
($0.1\la (r-R_s)/R_s\la 0.6$). The leak of particles into the low-$\psi$
region ($\psi\sim \psi_{\rm ISM}$) is less significant for $pc=20$~TeV
than for $pc=1$~TeV (Figure~\ref{fig:evo}b). This is because $\psi_{\rm
ISM}$ at $pc=1$~TeV is smaller than that at $pc=20$~TeV
(Figure~\ref{fig:evo}a and equations~\ref{eq:kappa_ism}
and~\ref{eq:psi_ism}).  For smaller $\psi_{\rm ISM}$, particles tend to
go father away from the shock front, if the distance is represented in
the units of $\kappa_{\rm B0}/V_s$. Thus, for $pc=1$~TeV, particles
escaping into the instep make the gradient of $f$ in the instep, while
for $pc=20$~TeV most particles are confined in the shin and make the
gradient of $f$ in the shin. The gradient of $f$ increases $\psi$
(equation~[\ref{eq:psi}]).

\begin{figure}
\includegraphics[width=84mm]{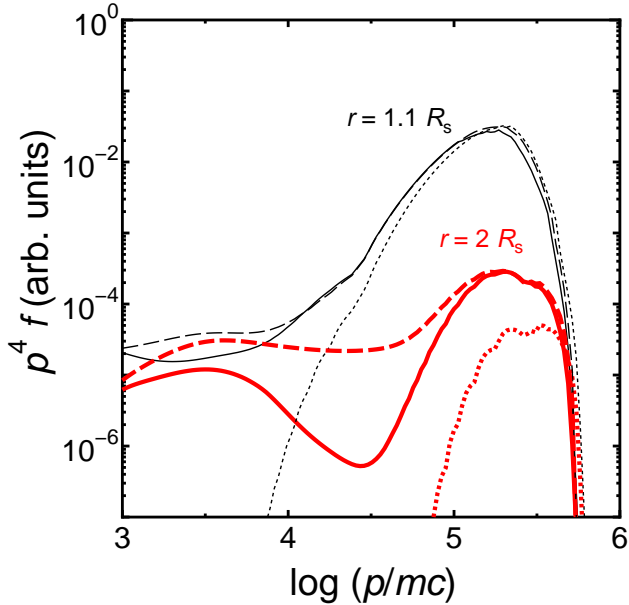}
\caption{Spectra of CRs at $r=1.1\: R_s$
 and $2\: R_s$ at $t=10\: t_0$. Solid, dotted, and dashed lines are for
 Models~A, B, and C, respectively. \label{fig:spr}}
\end{figure}

CR particles that escape from an SNR illuminate molecular clouds around
the SNR and could be observed in $\gamma$-rays through pion
production. Figure~\ref{fig:spr} shows the spectra of particles at
$r=1.1\: R_s$ and $2\: R_s$ at $t=10\: t_0$. In Model~B, only particles
with high energies can reach $r\sim 1.1\: R_s$, because the diffusion
coefficient for low-energy particles is small. On the other hand, in
Models~A and~C, even low-energy particles can reach this radius, and the
trend is more prominent at larger radii ($r\sim 2\: R_s$). Comparison
between Models~A and~C shows that the effect of the wave growth cannot
be ignored in the region away from the shock front ($r=2\: R_s$); the
energy density of particles at $p/mc \sim 10^{4.5}$ in Model~A is
smaller than that in Model~C by more than a factor of 10. In
Figure~\ref{fig:spr}, the broad hump of $p^4 f$ at $p/mc \sim 10^{3.5}$
at $r\sim 2\: R_s$ for Models~A and~C reflects the escape of particles
into the instep region of $\psi\sim\psi_{\rm ISM}$ shown in
Figure~\ref{fig:Bf} for $pc=1$~TeV. The dip at $p/mc \sim 10^{4.5}$ at
$r\sim 2\: R_s$ for Model~A is made because the particles hardly reach
that radius because of the growth of $\psi$ in the shin ($pc=20$~TeV in
Figure~\ref{fig:Bf}). The peaks at $p/mc \sim 10^{5.3}$ for Models~A, B
and~C mean that the region of $\psi\sim\psi_{\rm B}$ and the particles
confined in that region reach $r\sim 2\: R_s$ and those particles are
freely escaping in our treatment effectively. The decrease of spectra
for Models~A and~C at $r=2\: R_s$ for $p/mc \la 10^{3.5}$ suggests
that the spectra are affected by the outer boundary condition of $f=0$,
because $r_{\rm out}(p)$ is relatively close to $2\: R_s$. If $r_{\rm
out}(p)$ is set at a larger radius, we expect that more particles reach
$r\ga 2\: R_s$.

At $t=10\: t_0$, the total energies of CRs for $r>2\: R_s$ are
$4.0\times 10^{48}$, $2.3\times 10^{47}$, and $3.0\times 10^{49}$~erg
for Models~A, B and C, respectively. These results show that the growth
of the waves can significantly change the CR spectrum and density around
a middle-aged SNR, which should affect the $\gamma$-ray emission
originated from the CRs. In this study, we did not treat particles with
$pc\sim$~GeV. However, above discussion suggests that many of them can
escape into the instep region, because $\psi_{\rm ISM}$ for these
energies are fairly small. It is to be noted that the results shown here
do not involve the effect of wave damping. For example, if the ISM is
not fully ionised, the Alfv\'en waves may damp through collisions
between charged and neutral particles \citep{kul71,dru96}. If this
happens, the spectrum of the escaped CR particles would be significantly
changed \citep{fuj10}.

\section{Conclusion}

We have investigated the escape of CR protons accelerated at an SNR and
their diffusion in the surrounding ISM. We solved a diffusion-convection
equation from the vicinity of the shock front to the region far away
from the front. We also considered the amplifications of Alfv\'{e}n
waves generated by the escaping CR particles, which affects the
diffusion of the particles into the ambient ISM. We found that the
amplification of the waves reduces the diffusion coefficient on a scale
of the SNR and the escape of the particles is significantly delayed. Our
results suggests that the particle scattering by the waves should have
influence on the energy spectrum of CR particles observed at molecular
clouds around SNRs in the $\gamma$-ray band.

\section*{Acknowledgments}

We thank T.~Tsuribe for useful discussions. This work was supported by
KAKENHI (Y.~F.: 20540269; F.~T.; 20540231, K.~I.; 21-1979).

\end{document}